# Reversible Superconductivity in Electrochromic Indium-Tin Oxide Films

Ali E. Aliev,[*,1] Ka Xiong,[2] Kyeongjae Cho[2,3] and M. B. Salamon[4]

[1] *Alan G. MacDiarmid NanoTech Institute, University of Texas at Dallas, Richardson, TX, 75083*
[2] *Department of Materials Science and Engineering, University of Texas at Dallas, Richardson, TX, 75083*
[3] *WCU Multiscale Mechanical Design Division, School of Mechanical and Aerospace Engineering, Seoul National University, Seoul 151-742, Republic of Korea*
[4] *Department of Physics, University of Texas at Dallas, Richardson, TX, 75083*

**Abstract**

Transparent conductive indium tin oxide (ITO) thin films, electrochemically intercalated with sodium or other cations, show tunable superconducting transitions with a maximum $T_c$ at 5 K. The transition temperature and the density of states, $D(E_F)$ (extracted from the measured Pauli susceptibility $\chi^p$) exhibit the same dome shaped behavior as a function of electron density. Optimally intercalated samples have an upper critical field ≈ 4 T and $\Delta/k_B T_c \approx 2.0$. Accompanying the development of superconductivity, the films show a reversible electrochromic change from transparent to colored and are partially transparent (orange) at the peak of the superconducting dome. This reversible intercalation of alkali and alkali earth ions into thin ITO films opens diverse opportunities for tunable, optically transparent superconductors.

All superconductors revert to their normal-state optical properties for light whose photon energy significantly exceeds twice the energy gap, and most are consequently either metallic-shiny or opaque. However, many applications have been proposed[1] (and even patented[2]) that require only the development of practical, optically transparent superconductors. We report here a very simple means of creating a *reversible* superconducting regime by electrochemical intercalation (and de-intercalation) of sodium (and other cations) into commercially available indium-tin-oxide (ITO) films, typically ($In_{2-y} Sn_y O_{3-\delta}$, y~0.2). As is characteristic of many compound superconductors, doped ITO exhibits a superconducting (SC) "dome", in which the transition temperature increases to a maximum and then decreases with changes in a "control parameter". Perhaps the best known example is the SC dome in $YBa_2Cu_3O_x$, where the oxygen content controls the SC transition temperature $T_c$.[3,4] In Fe pnictides, substitution of Co for Fe induces the formation of a SC dome;[5] many heavy Fermion materials exhibit similar effects as functions of hydrostatic pressure.[6] In the ITO case, alkali-metal content is the control parameter and the maximum achievable transition temperature is 5 K, well above both In (3.4 K) and Sn (3.7 K). We are able to correlate the SC dome, the density of states at the Fermi surface and the electrochromic change from a transparent to a partly transparent (colored) film with the intercalation-induced increase in electron density. While we focus here on electrochemical doping with sodium; similar results will be reported elsewhere for other alkalis, alkali earth, or ammonium cations, using both aqueous and non-aqueous (propylene carbonate, PC) electrolytes.

---

[*] Corresponding author. Tel.: +1 972-883-6543; fax: +1 972-883-6529. *E-mail address:* Ali.Aliev@utdallas.edu.

The substitution of Sn for In in $In_2O_3$ introduces a hybridization gap,[7] pushing the valence band below the Fermi energy. A plasma frequency in the infrared and a partially filled conduction band together give ITO its characteristic optical transparency and high conductivity. The crystal structure (bixbyite) of the parent compound contains 16 formula units per unit cell. The oxygen atoms occupy 48 crystallographically equivalent sites, leaving an ordered array of 16 vacant sites per unit cell. The combination of vacancies and a large interatomic spacing make ITO favorable for intercalation by small ions. Even the parent compound, $In_2O_3$, in the form of an amorphous film, exhibits superconductivity with $T_c < 0.3$ K when the carrier density exceeds $10^{20}$ cm$^{-3}$ [8] and up to 3 K when combined with ZnO.[9] Further, Mori [10] reported a sharp SC transition in post-oxidized ITO films, with $T_c$ in the range 2-4 K. These results suggest that $T_c$ can be enhanced by increasing the carrier density via chemical doping.

Fig. 1 shows the X-ray diffraction pattern from an as-received ITO film from LumTech, Corp. (G002, film thickness $d$=400 nm). The texture is predominantly (100), the facets of which are the most open for intercalation. Samples are charged for varying periods of time at a current of $I$=0.2 mA in 1 M NaCl in $H_2O$. After intercalation, samples were dried in air and then annealed at 150-170 °C for one hour. The total Na content of the film is subsequently determined by energy dispersive X-ray spectrometry (EDS) analysis and spans the range 0-4 ±0.2 atomic % (at.%). Charged samples are stable, although some oxidation was observed in over-doped samples prepared in aqueous solutions for x > 3 at.%. No such effect was observed when PC was used as the electrolyte. It will be shown below that only a fraction of the deposited Na is actually intercalated into the ITO film.

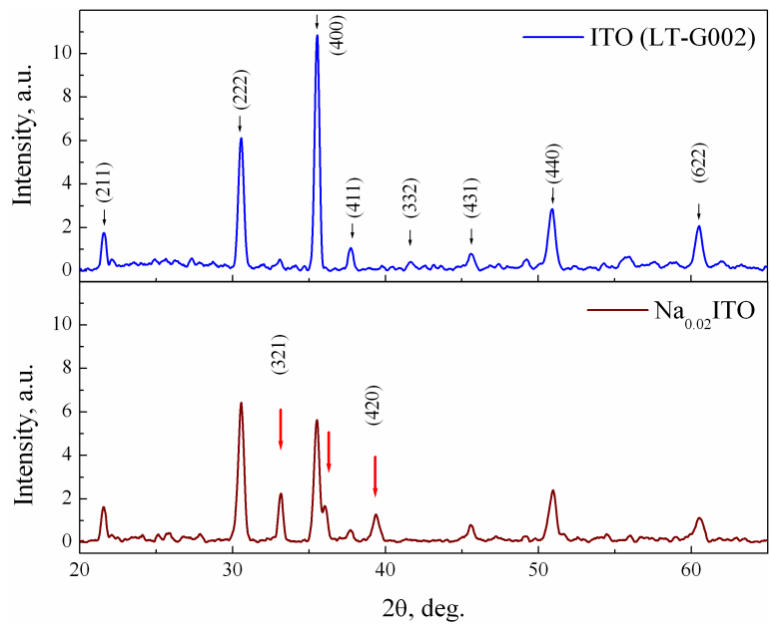

FIG. 1. X-ray diffraction pattern of bare ITO film (LT) and ITO film intercalated with Na$^+$ ions. Peaks marked by red arrows in lower panel are the new reflections peculiar to the Na$_x$ITO composite.

The lower part of Fig. 1 shows that the relative intensities of (222) and (400) reflections change while (321) and (420) peaks, which are weak in ITO, strengthen dramatically at a nominal EDS concentration of 2 at.%. The intensities change with charging time, providing strong evidence that the intercalated Na is incorporated into the underlying ITO crystal structure, possibly into vacant 16c sites. An unexpected peak develops just above the (400) position ($2\theta=36.2°$) which becomes more prominent with increased charging. The simulation described below confirms that all three peaks result from Na ions intercalated into vacant 16c sites. There is no substantial change in the position of the bixbyite Bragg peaks, indicating that intercalation occurs without significantly increasing the volume of the unit cell.

To gain insight at the atomic level the mechanisms of Na intercalation in ITO, we use first principles calculations to investigate atomic structures and the corresponding XRD patterns of Na-doped ITO. The atomic structures of Na-doped ITO are calculated by total energy plane-wave basis code VASP. The corresponding XRD patterns are then calculated by the REFLEX module of Materials Studio software package. For the simulation model, we use one unit cell of the bixbyite $In_2O_3$ that contains 80 atoms. In our model one In atom is replaced by Sn. By comparing the calculated total energies it is found that Sn prefers to occupy the 8b site. Since there are vacant 16 oxygen (16c) sites in $In_2O_3$, we place one or two Na at various sites and analyze the most stable atomic configuration. The relaxed atomic structures reveal that, indeed, it is energetically favorable for Na to occupy the vacant O site. It is found that the distance between the inserted Na and nearest In is 2.6 Å. Further, we can confirm that Na acts as an electron donor in ITO shifting the Fermi level to +0.4 eV suggesting a rigid band doping model. More importantly, the change of the XRD patterns between undoped ITO and Na-doped ITO is in excellent agreement with the experimental data. The supplementary material shows a comparison of calculated XRD patterns of Sn-doped $In_2O_3$, one Na per unit cell (16 formula units) of $In_2O_3$ and two Na per $In_2O_3$ unit cell. As found experimentally, the intensity of the (321) and (411) peaks increases sharply as the Na concentration increases, more than doubling for two Na per unit cell. Meanwhile, we observe that the intensity of the (400) peak decreases slightly as the number of the incorporated Na ions increases.

Fig. 2(a) shows the development of the SC dome as a function of charging time. In addition to the appearance of superconductivity, other properties of ITO films change dramatically upon intercalation. The films change from transparent to dark brown with increasing charge. While commercial ITO films on glass substrates exhibit semiconducting behavior below 100 K, the intercalated films are metallic ($dR/dT >0$, but small) over the full temperature range. The room temperature resistance, however, increases by two orders of magnitude. The highest $T_c$, with an onset above 5 K, occurs at a charging time near 1000 s at 0.1 mA/cm$^2$, but then decreases with increasing charging. To determine the true intercalation level, we have performed Hall-effect measurements in a 1 T field over the temperature range 100-200 K. Before charging, the $n$-type carrier density in our ITO samples is $9.2 \times 10^{20}$ cm$^{-3}$, increasing to $1.2 \times 10^{21}$ cm$^{-3}$ after charging for 5000 s. These densities are consistent with 3 at.% Sn substitution for In (one Sn per ITO unit cell) and plus 0.3 electrons per cell added by intercalation.

In Fig 2(b), we plot the transition temperature (open circles), defined at $R = 0.9R_n$, vs the Hall-derived carrier density. A maximum occurs at $n = 1.05 \times 10^{21}$ cm$^{-3}$, consistent with the carrier densities at which superconductivity appears in oxygen-deficient ITO.[8] The colored bar indicates the change in the appearance of the film, passing from transparent to yellow-orange to brown and finally to black with increasing Na content. Samples can be bleached (and superconductivity suppressed) upon discharging under the same electrolytic conditions. Note that the superconducting $T_c$ can be predicted from the room temperature color of the film.

To make contact with the BCS theory, we have measured the normal state Pauli susceptibility to extract the density of states at the Fermi level, $D(E_F)$. We subtract from the data a Curie ($1/T$) and diamagnetic background contributions from the glass substrate as well as the Langevin core diamagnetism of the ITO film. The remaining temperature independent paramagnetic signal is attributed to the free-electron Pauli susceptibility $\chi^p = 2\mu_B^2 D(E_F)$ from which $D(E_F)$ was estimated. These are plotted as a solid circles curve in Fig. 2(b). Note that $D(E_F)$ has a maximum at the carrier density at the top of the SC dome which we attribute to filling of the conduction band toward the hybridization gap. The density of states per cell in ITO (comprising 16 formula units) is consistent with calculations using density-functional theory [7, 11] and an experimental study of Sn-doped In$_2$O$_3$ (2 and 10 weight% of Sn) using hard X-ray photoemission spectroscopy (XPS).[12] Indeed, those calculations predict a $D(E_F)$ maximum approximately 0.2 eV above the Fermi level of ITO; accommodating 0.3 states/cell in a rigid band picture and with DOS in the range of 2-3 states/eV cell requires a comparable shift in Fermi level by approximately 0.1-0.15 eV. As further evidence, XPS data, obtained using a Perkin Elmer ESCA System, PHI 5600 confirm that additional electrons are injected by intercalation and populate the new band originated mainly from mixing of Sn 5s and O 2p states. The onset of this XPS peak in the conduction band is located at a Fermi energy $E_F$= 2.7±0.2 eV above the valence band, consistent with the optical transmission onset developed during coloration.

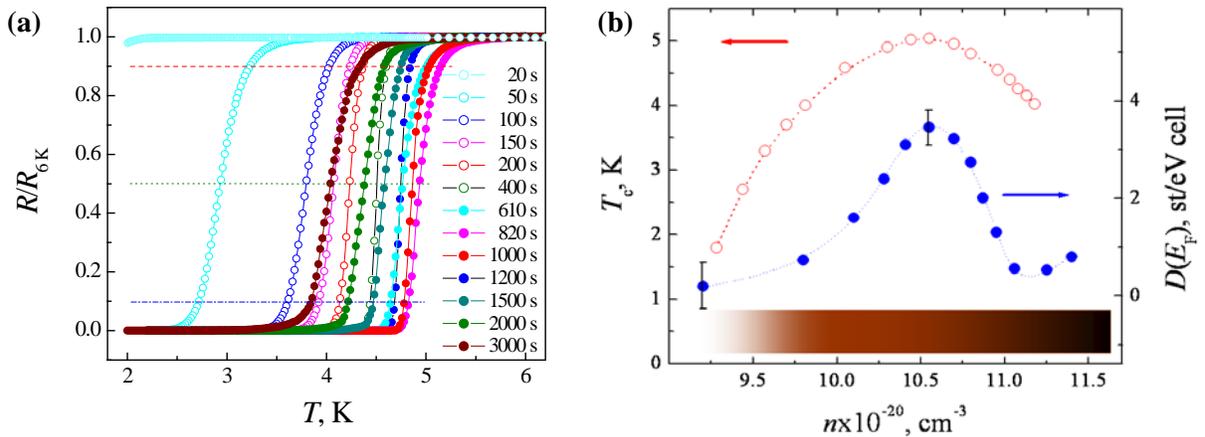

FIG 2. (a) The temperature dependence of normalized ac resistance ($I_{exc}$=1 μA) for ITO film intercalated in aqueous 1M NaCl electrolyte at fixed current of 0.1 mA/cm$^2$ for intercalation time increased from 20 s to 3000 s. (b) The concentration dependences of $T_c$ taken at resistance onset level of 0.9 $R_n$ (open circles), and $D(E_F)$ extracted from the Pauli susceptibility measurements (solid circles). The horizontal bar shows the development of ITO film's color during intercalation.

The inset to Fig. 3 shows the dependence of the SC transition for $Na_{0.02}ITO$ ($Q$=0.1 mA/cm$^2$ x 1000 s) on magnetic field. With magnetic field aligned both perpendicular and parallel to the film plane, the upper critical field $H_{c2}(T)$, defined at the transition onset at $0.9R_n$, is linear in temperature (Fig. 3). The relatively high critical fields, $H_{cr}(0) \approx 4$ T in sodium- (and magnesium) doped ITO films, correspond to a short Ginzburg-Landau coherence length. However, the resistivity of Na-doped ITO is $\rho$=2.5 mΩ·cm from which we obtain a mean-free path $\ell_{tr} = \hbar(3\pi^2)^{1/3}/n^{2/3}e^2\rho = 0.5$ nm for nearly-free electrons with a density $n = 1.05 \times 10^{21}$ cm$^{-3}$. From the well-known WHH expression [13]

$$\xi_{GL}^2(0) = \left(\frac{\phi_0}{2\pi T_c}\right) \cdot \left[(\frac{dH_{c2}}{dT})_{T=T_c}\right]^{-1}, \quad (1)$$

we find the Ginzburg-Landau length $\xi_{GL}$ = 9 nm, which is close to the value deduced from the linear $H_{c2}(T)$ for 10 nm thick In/In$_2$O$_x$ layered nanocomposites,[14] ($\xi(0)$=6 nm) and for 20 nm thick annealed ITO film on polyester,[9] ($\xi(0)$ =7.7 nm). In the dirty limit, where $\xi_{GL} = 0.852(\xi_{BCS}(0)\cdot\ell_{tr})^{1/2}$, this gives the BCS coherence length $\xi_{BCS}(0)$ = 224 nm. For a Fermi velocity $v_F = 9 \times 10^5$ m/s (see below) we obtain the energy gap $\Delta(0) = 2.0 k_B T_c$. It appears that Na-doped ITO is a dirty, type II superconductor in the strong-coupling regime. Assuming a Debye temperature $\theta_D \approx 1000$ K,[15] we find the effective strong-coupling parameter is $\lambda_{eff} = \{\ln(\theta_D/T_c)\}^{-1} \approx 0.2$.

Additional support for free-electron-like behavior comes from our measurement of the Seebeck coefficient $S$, which is negative and linear in temperature down to $T_c$. The diffusive thermopower at temperatures well below $\theta_D \approx 1000$ K is given by the Mott expression when doped impurities rather than acoustic phonons dominate electron scattering ($r = 0$),[16]

$$S = \frac{\pi^2}{3}\frac{k_B}{|e|}\left(r+\frac{3}{2}\right)\left(\frac{k_b T}{E_F}\right) = \frac{\pi^2}{2}\frac{k_B^2}{|e|}\frac{T}{E_F}. \quad (2)$$

as is the case for intercalated samples. The linear slope of $S(T)$ in the normal state corresponds to a Fermi energy of $E_F$=1.05 eV and is consistent with an electron density of $1.2 \times 10^{21}$ cm$^{-3}$.

A large number of ITO films from different sources, deposited under different conditions, have been investigated. Details of their superconducting behavior and further results, including detailed XPS studies, behavior near $T_c$ and data on the intercalation of other cations, will be presented elsewhere.

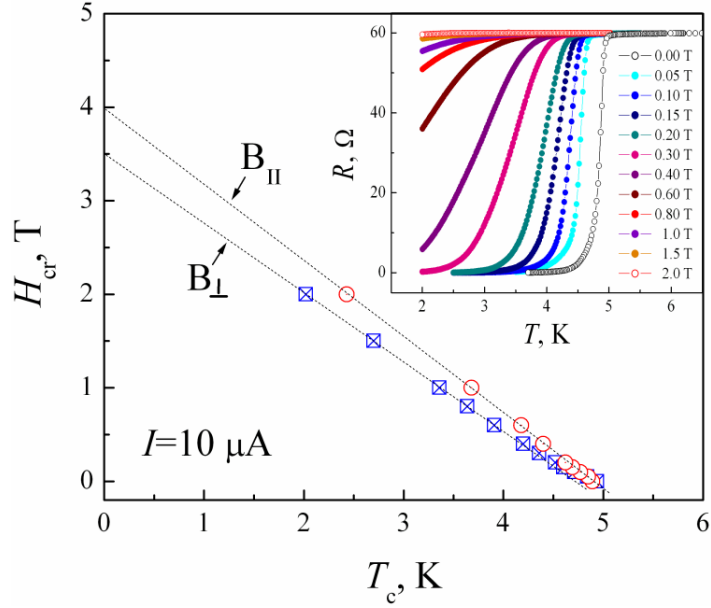

FIG. 3. The temperature dependence of upper critical magnetic field $H_{c2}$ measured for magnetic field applied perpendicular (open squares), and parallel to the film (open circles) in $Na_{0.02}ITO$ ($Q$=0.1 mA/cm$^2$ x 1000 s). The inset shows the temperature dependence of film resistance at increased external magnetic fields applied perpendicular to the film.

There are numerous examples in the literature of superconductivity induced by intercalation, $C_6Ca$ and relatives [17] being recent cases. Electrochemical methods have also been used to intercalate Li into various perovskites.[18] However, the current work is the first to demonstrate the continuous, reversible evolution of the superconducting $T_c$ with doping and correlate to both the density of states and the change in sample color from transparent through orange to black. As the electron density increases from $n = 9 \times 10^{20}$ cm$^{-3}$ (pure ITO) to $1.2 \times 10^{21}$ cm$^{-3}$, the superconducting onset (10% drop in resistivity) reaches 5 K when $n = 1.05 \times 10^{21}$ cm$^{-3}$. At the top of the dome, a small resistivity decline starts at 9 K, while the magnetic susceptibility measurements exhibit a ZFC onset at 4.7 K, at which point the resistivity curve reaches its lowest measurable level (<10$^{-5}$ Ω). Our results are summarized in Table I.

| $T_c$, K | $n$, 10$^{21}$ cm$^{-3}$ | $\xi_{GL}$, nm | $l_{tr}$, nm | $\xi_0$, nm | $v_F$, 10$^5$ m/s | $\Delta(0)/k_B T_c$ |
|---|---|---|---|---|---|---|
| 5 | 1.05 | 9 | 0.5 | 224 | 9.4 | 2.0 |

Table I. Parameters deduced for optimally intercalated Na$_x$ITO. We assume an effective mass of $m^*$=0.4 $m_e$,[19] a spherical Fermi surface with the Fermi energy $E_F$ = 1.05 eV determined from the Seebeck coefficient, and an extrapolated upper critical field of 4 T. The Fermi velocity is determined from $v_F = (2E_F/m^*)^{1/2}$. The same value follows from the Hall density of carriers n, $v_F = \hbar(3\pi^2 n)^{1/3}/m^*$.

A remarkable aspect of the current work is that the Na ions enter the underlying bixbyite lattice with no change in crystal structure and only very small changes in lattice constant. We have suggested that the additional cations occupy vacant 16c sites in the structure, changing only the relative intensities of allowed x-ray lines. Complete structural analysis will be described later. The results reported here are not unique to Na intercalation, and future work will show similar results for other small cations, even doubly-charged ones. This technique opens pathways to the discovery of new superconducting phases, employing other conducting oxides.

We acknowledge helpful discussions with I. Mazin. This work was partially supported by the Air Force Office of Scientific Research grant FA9550-09-1-0384. K.C. was supported by the NRF of Korea through WCU program (Grant No. R-31-2009-000-10083-0).

# Reversible Superconductivity in Electrochromic Indium-Tin Oxide Films
## Supplementary material

Ali E. Aliev,[*,1] Ka Xiong,[2] Kyeongjae Cho[2,3] and M. B. Salamon[4]

[1] *Alan G. MacDiarmid NanoTech Institute, University of Texas at Dallas, Richardson, TX, 75083*
[2] *Department of Materials Science and Engineering, University of Texas at Dallas, Richardson, TX, 75083*
[3] *WCU Multiscale Mechanical Design Division, School of Mechanical and Aerospace Engineering, Seoul National University, Seoul 151-742, Republic of Korea*
[4] *Department of Physics, University of Texas at Dallas, Richardson, TX, 75083*

*S1. The XRD patterns simulated using REFLEX module of Materials Studio software package.*

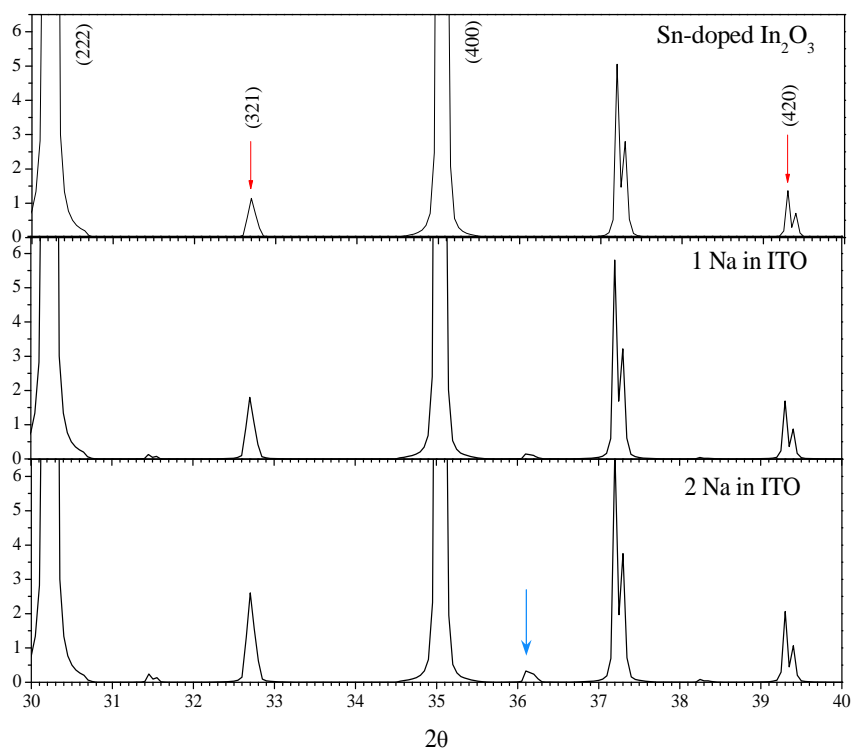

Fig. S1. A comparison of calculated XRD patterns of Sn-doped $In_2O_3$, one Na in Sn-$In_2O_3$, and two Na in Sn-$In_2O_3$. In addition to (321) and (420) peaks enforced by insertion of Na ions in vacant oxygen 16c sites an unknown peak at 36.2 degree is appeared (blue arrow). This peak, perhaps with lower space group symmetry (410), creates the overlapping shoulder on (400) peak in experimentally obtained XRD.

[*] Corresponding author. Tel.: +1 972-883-6543; fax: +1 972-883-6529. *E-mail address:* Ali.Aliev@utdallas.edu.

*S2. The atomic structures of Sn-doped In$_2$O$_3$ calculated by total energy plane-wave basis code VASP.*

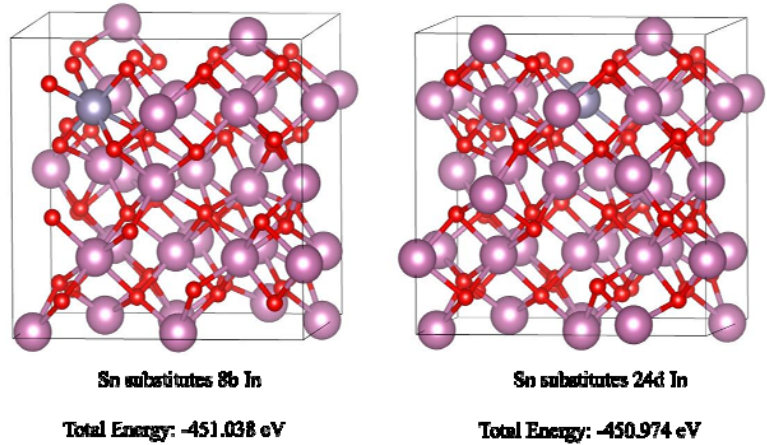

Fig.S2. The Sn substitution of In atoms in 8b positions are more favorable.

*S3. The atomic structures of Na intercalated ITO calculated by total energy plane-wave basis code VASP.*

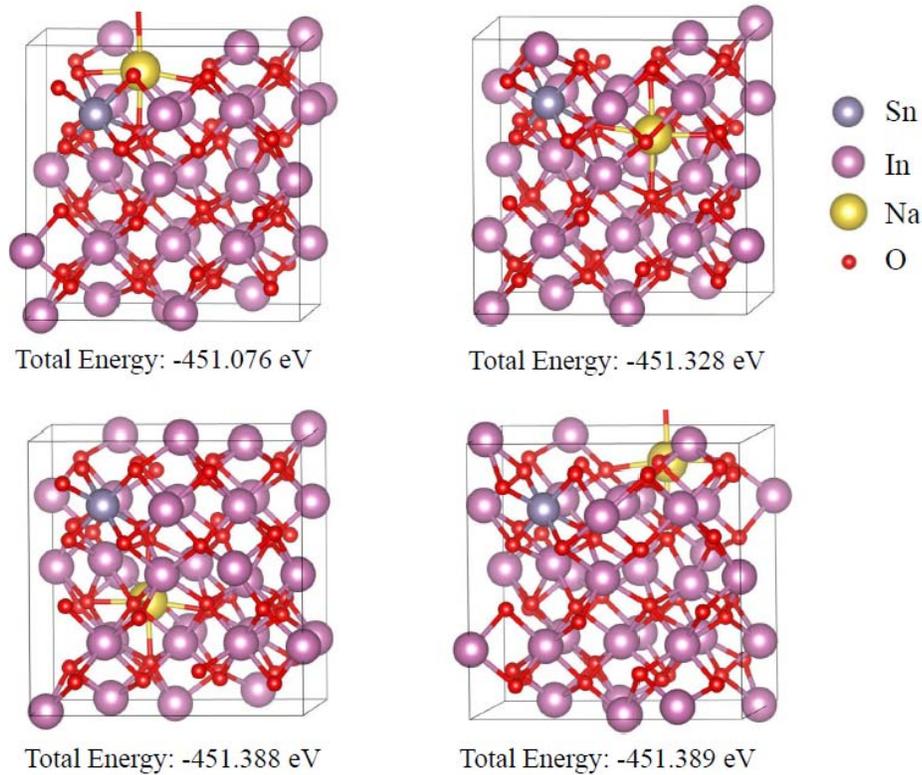

Fig. S3. The relaxed atomic structures reveal that, it is energetically favorable for Na to occupy the vacant (16c) O site and staying away from the Sn rather than stay at the nearest neighbor site.

*S4. The calculated density of states of undoped and Na doped In$_2$O$_3$, and Sn doped In$_2$O$_3$ (Sn:In$_2$O$_3$) and Na in Sn:In$_2$O$_3$ calculated by total energy plane-wave basis code VASP.*

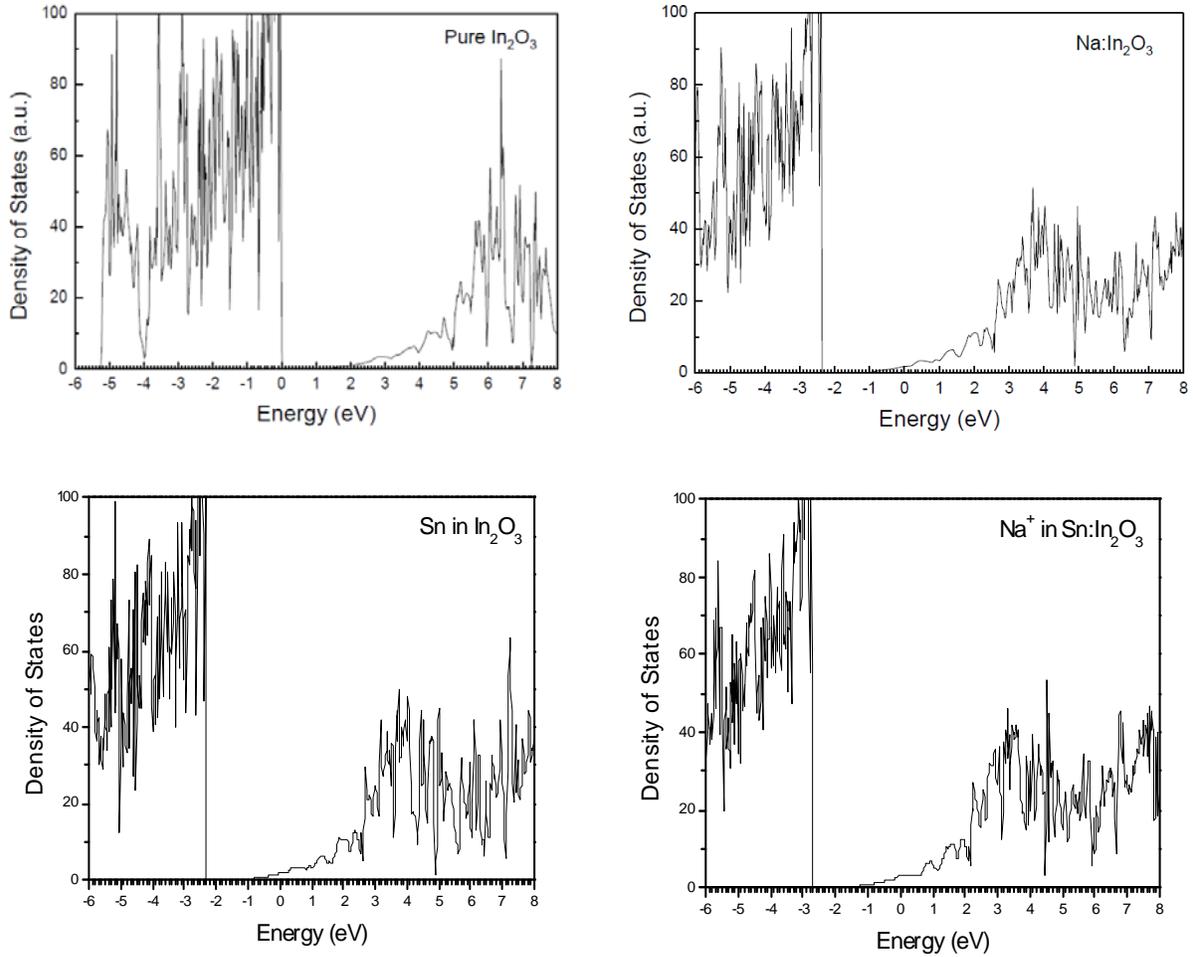

Fig. S4. A comparison of the calculated total density of states of undoped In$_2$O$_3$, Na doped In$_2$O$_3$, Sn doped In$_2$O$_3$ (Sn:In$_2$O$_3$) and Na in Sn:In$_2$O$_3$. The Fermi level is located at 0 eV. It is shown that both Na and Sn act as donors in In$_2$O$_3$, where the Fermi level is located inside the In$_2$O$_3$ conduction band. For Na in Sn:In$_2$O$_3$, the Fermi level also lies inside the In$_2$O$_3$ conduction band, making the oxide n-type. The sodium insertion shifts the Fermi level to 0.4 eV but does not change the DOS shape substantially.